\documentclass[aps,prd,preprint,tightenlines,superscriptaddress,nofootinbib,eqsecnum,amsmath,amssymb]{revtex4}
\usepackage{graphicx}
\usepackage{multirow} 

\makeatletter
\def\frontmatter@title@above{\vspace*{1.5\baselineskip}}%
\def\frontmatter@title@format{\LARGE\bfseries\centering\parskip\z@skip}%
\def\frontmatter@title@below{\addvspace{\baselineskip}}%
\def\frontmatter@preabstractspace{2.5\baselineskip}

\def\frontmatter@postabstractspace{1.5\baselineskip}
\makeatother

\newcommand{\ie}{\textit{i.e.}}
\newcommand{\program}[1]{\texttt{#1}}

\begin{document}

\preprint{FTUV-11-0715}
\preprint{IFUM-972-FT}
\preprint{KA-TP-16-2011}
\preprint{LPN11-38}
\preprint{SFB/CPP-11-39}

%Title of paper
\title{$Z\gamma\gamma$ production with leptonic decays and triple photon
        production at NLO QCD}

\author{G. Bozzi}
\affiliation{Dipartimento di Fisica, Universit\`a di Milano and INFN, Sezione di Milano, \\
     20133 Milano, Italy}
\author{F. Campanario}
\affiliation{Institut f\"ur Theoretische Physik, Universit\"at Karlsruhe, \\
  Karlsruhe Institute of Technology, 76128 Karlsruhe, Germany}
\author{M. Rauch}
\affiliation{Institut f\"ur Theoretische Physik, Universit\"at Karlsruhe, \\
  Karlsruhe Institute of Technology, 76128 Karlsruhe, Germany}
\author{D. Zeppenfeld}
\affiliation{Institut f\"ur Theoretische Physik, Universit\"at Karlsruhe, \\
  Karlsruhe Institute of Technology, 76128 Karlsruhe, Germany}

\date{\today}

%%%%%%%%%%%%%%%%%%%%%%%% abstract %%%%%%%%%%%%%%%%%%%%%%%%%%%%%%%%%%%%%%%
\begin{abstract}
We present a calculation of the $\mathcal{O}(\alpha_s)$ QCD corrections
to the production of a $Z$ boson in association with two photons and to
triple photon production at hadron colliders. All final-state photons
are taken as real. For the $Z$ boson, we consider the decays both into
charged leptons and into neutrinos including all off-shell effects.
Numerical results are obtained via a Monte Carlo program based on the
structure of the \program{VBFNLO} program package. This allows us to
implement general cuts and distributions of the final-state particles.
We find that the NLO QCD corrections are sizable and significantly
exceed the expectations from a scale variation of the leading-order
result. In addition, differential distributions of important observables
change considerably. The prediction of two-photon-associated $Z$
production with $Z$ decays into neutrinos from the charged-lepton rate
works well, once we use an additional cut on the invariant mass of the
charged-lepton pair.
\end{abstract}

% insert suggested PACS numbers in braces on next line
%\pacs{}
% insert suggested keywords - APS authors don't need to do this
%\keywords{}

%\maketitle must follow title, authors, abstract, \pacs, and \keywords
\maketitle

%%%%%%%%%%%%%%%%%%%%%%%% body %%%%%%%%%%%%%%%%%%%%%%%%%%%%%%%%%%%%%%%%%%%
\section{Introduction}

The Large Hadron Collider (LHC) at CERN allows us to extend new-physics
searches into energy regions never tested before. To claim a discovery of
new phenomena, or provide improved bounds on them, it is important to
know accurate predictions on their Standard Model (SM) background. This
forces us to perform calculations beyond leading order, for integrated
cross sections as well as for differential distributions.
In this paper we present a calculation of the next-to-leading-order
(NLO) QCD corrections to $Z$-boson production in association with two
photons including off-shell effects for two decay modes
\begin{align}
\raisebox{-0.5\baselineskip}[0pt][0pt]{$``Z\gamma\gamma"\quad \biggl\{$} \quad
``Z_\ell\gamma\gamma" \qquad 
pp, p\bar{p} \rightarrow Z \gamma\gamma  + X &\rightarrow \ell^+ \ell^-
\gamma\gamma + X  
  \label{eq:ZellAA}\\
``Z_\nu\gamma\gamma" \qquad 
pp, p\bar{p} \rightarrow Z \gamma\gamma + X  
 &\rightarrow \bar{\nu} \nu \gamma \gamma  + X \label{eq:ZnuAA} \,,
\\
\intertext{and to triple photon production}
``\gamma\gamma\gamma" \qquad 
pp, p\bar{p} \rightarrow \gamma \gamma\gamma  + X& \label{eq:AAA} \,.
\end{align}
Both processes provide backgrounds for new-physics searches (for an
overview, see e.g.  Ref.~\cite{Campbell:2006wx}). The process
(\ref{eq:ZnuAA}) with its signature of two photons and missing
transverse energy appears for example as background in 
models with gauge-mediated supersymmetry breaking (GMSB)~\cite{GMSB}. 
In such models, the next-to-lightest supersymmetric particle is often a
bino-like neutralino, which will decay into a gravitino and a photon.
Since supersymmetric particles are pair-produced due to R-parity
conservation, this leads exactly to the signature studied here.
Triple-photon production provides a background to techni-pion production
in association with a photon~\cite{Zerwekh:2002ex}, where the
techni-pion decays into a photon pair.

With this calculation the determination of triple vector-boson
production cross sections at NLO QCD precision at hadron colliders is
complete. The other processes belonging to this class have already been
computed in
Refs.~\cite{Lazopoulos:2007ix, Hankele:2007sb, Campanario:2008yg,
Binoth:2008kt, Bozzi:2009ig, Bozzi:2010sj, Baur:2010zf, Bozzi:2011ww}. 
Multi-vector-boson production with one additional jet has been studied
at NLO QCD so far for the diboson processes $W W j$, $W \gamma j$, 
$W Z j$ and $Z Z j$~\cite{Campbell:2007ev, Dittmaier:2009un,
Campanario:2009um, Campanario:2010hp, Binoth:2009wk} and the triboson
process $W \gamma \gamma j$~\cite{Campanario:2011ud}.

The outline of the paper is as follows: in Section~\ref{sec:calc} the
Feynman diagrams which appear in our calculation are shown. We discuss
the different techniques used to compute the virtual and real
corrections and give an overview of the various checks which have been
performed to ensure the correctness of the calculation. In
section~\ref{sec:numeric} numerical results are presented. This includes
scale variations of the leading-order (LO) and NLO integrated cross
sections as well as selected differential distributions. We compare
the decay of the $Z$ boson into neutrinos, where the photons can only be
radiated off the quark line, with the one into charged leptons, where
also final-state photon radiation off the leptons is possible. A cut on
the invariant mass of the lepton pair, restricting it to a small region
around the $Z$ mass, allows us to suppress the latter contributions.
Section~\ref{sec:concl} summarizes our work.

%We have performed our calculation within the framework of
%\program{VBFNLO}~\cite{Arnold:2008rz}, which is a flexible parton-level Monte
%Carlo program and allows the user to define general differential
%distributions and acceptance cuts.

\section{Calculational Details}
\label{sec:calc}
\begin{figure}[htbp!]
\begin{center}
\includegraphics[scale=0.5]{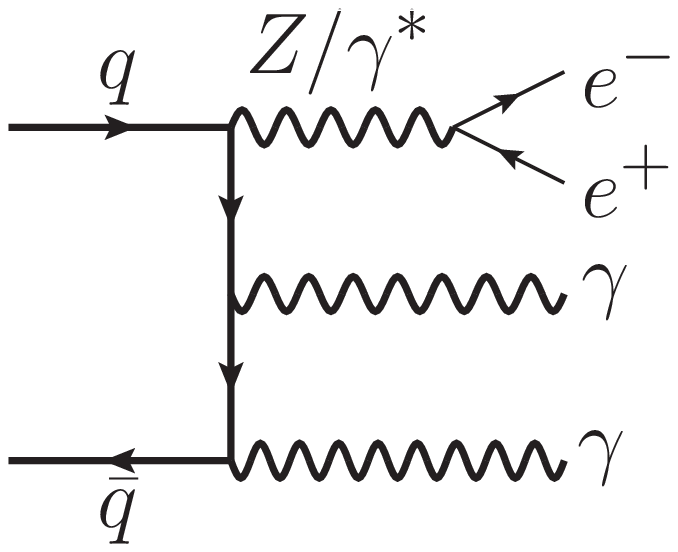}\hfill
\includegraphics[scale=0.5]{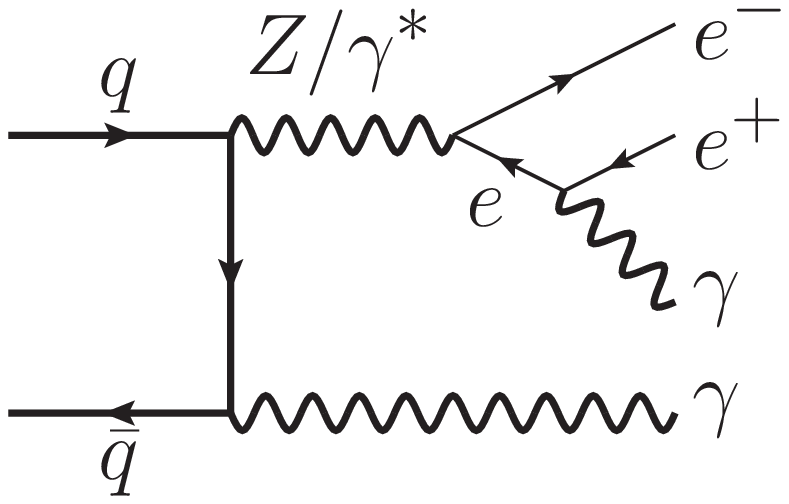}\hfill
\includegraphics[scale=0.5]{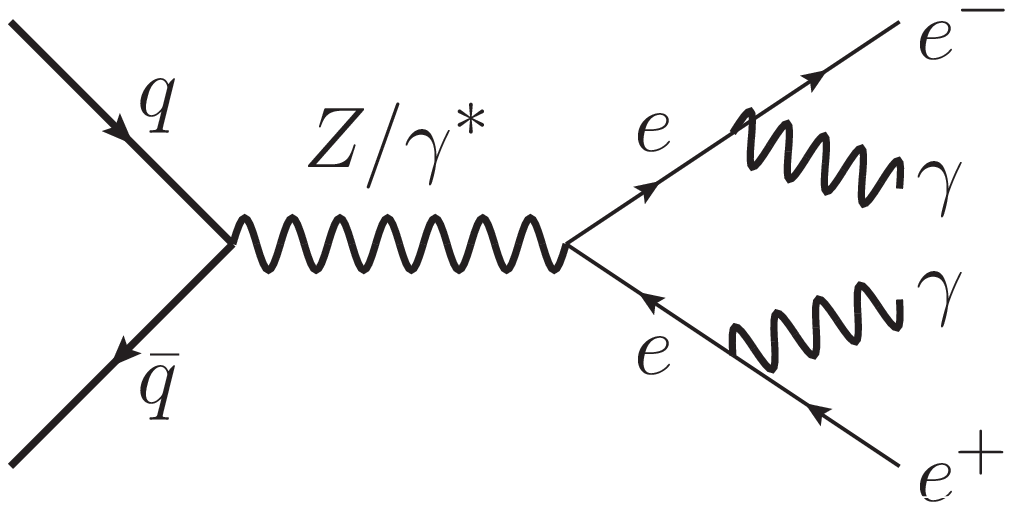} \\[\baselineskip]
\includegraphics[scale=0.5]{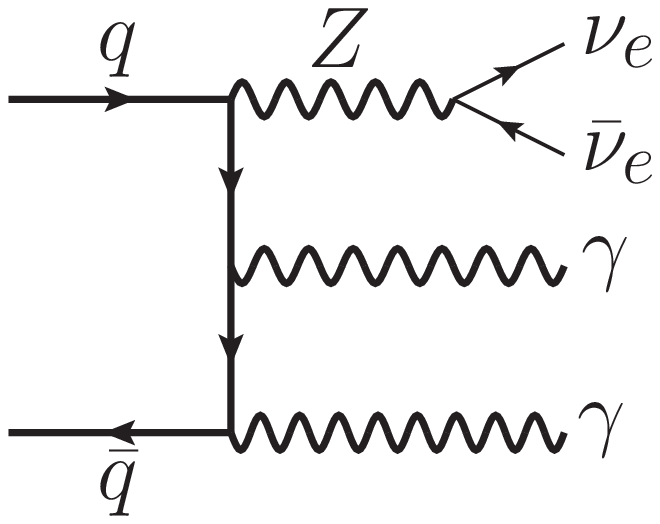}\qquad
\includegraphics[scale=0.5]{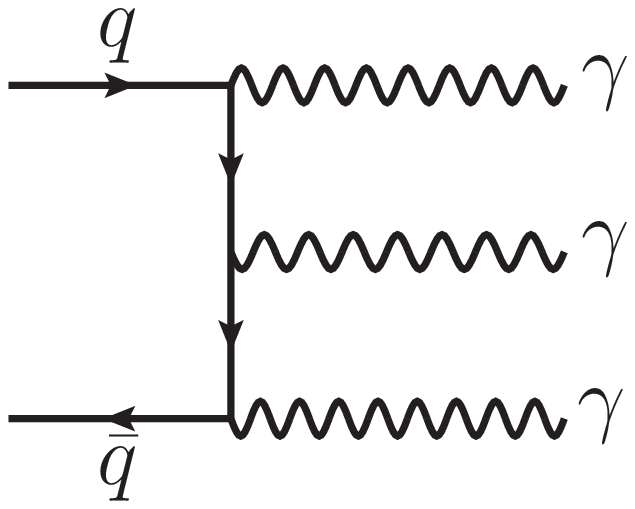}
\end{center}
\caption{Examples of topologies of Feynman diagrams
  contributing to the different processes at tree level.
  {\it Top row:} $pp\to Z\gamma\gamma$ + X including $Z$
  decays into a pair of charged leptons.
  {\it Bottom row:} $pp\to Z\gamma\gamma$ + X including $Z$
  decays into a pair of neutrinos (left) and 
  $pp\to \gamma\gamma\gamma$ + X (right). }
\label{fig:feyn}
\end{figure}
We calculate all contributions up to order $\alpha_s\alpha^4$ to the
processes~(\ref{eq:ZellAA}) and~(\ref{eq:ZnuAA}) in the limit of
massless fermions. In Fig.~\ref{fig:feyn}, we show examples of
Feynman diagrams appearing in our calculation. 
The different possibilities for the decay into charged leptons are
depicted in the top row. Either all three vector bosons can be attached
to the quark line (left), or one (middle) or both (right) photons are
radiated off the charged leptons. In total, 40 different Feynman
diagrams appear at leading order. To improve the speed of the
calculation, we use the technique of ``leptonic tensors'' as described
in Ref.~\cite{Jager:2006zc}. Invariant subparts of the amplitude, namely
the decays of the $Z$ and virtual photons, are computed independently of
the rest of the amplitude and only once per phase-space point. The
result can then be easily reused when calculating the matrix elements,
which greatly reduces the required computation time. 
Up to permutations of the vector bosons, only the diagram on the bottom
left of Fig.~\ref{fig:feyn} can appear at leading order for $Z$ decays
into a pair of neutrinos. Hence, the photons are always radiated off the
quark line.

The process~(\ref{eq:AAA}) is correspondingly calculated up to order
$\alpha_s\alpha^3$.  Up to permutations of the photons, only the Feynman
diagram depicted in the lower right of Fig.~\ref{fig:feyn} contributes
at leading order.

To compute all these matrix elements, we use the helicity formalism of
Ref.~\cite{Hagiwara:1985yu}. 
Contributions from virtual diagrams, which interfere with the
tree-level diagrams, and from real emission occur at NLO QCD. Both are
infrared divergent, but the sum of them does not contain any infrared
divergences according to the Kinoshita-Lee-Nauenberg (KLN)
theorem~\cite{Kinoshita:1962ur}.  We achieve this cancellation in a
numerically stable way by using the Catani-Seymour dipole subtraction
method \cite{Catani:1996vz}, which analytically extracts the divergences
which are then canceled against the virtual ones.  
Part of the initial-state collinear
singularities is factored into the parton-distribution functions, giving
rise to additional so-called ``finite collinear terms''.
The NLO real corrections are obtained from the LO diagrams by attaching
a gluon to the quark line in every possible way. This gluon can then be
either a final-state particle, radiated off the quark line, or an
initial-state one splitting into a quark--anti-quark pair. For the
$Z_\ell\gamma\gamma$ case there are 120 Feynman diagrams for each
possibility. Hence, using leptonic tensors is particularly useful in
this case.

We obtain the virtual NLO QCD corrections by inserting gluon lines into
the diagrams of Fig.~\ref{fig:feyn} in every possible way. This yields
loop diagrams with up to five external legs, \ie{} pentagon
diagrams. The occurring five-point integrals are solved using the
prescription given in Ref.~\cite{Denner:2002ii}, which does not need any
inverse Gram determinants, with the setup of
Ref.~\cite{Campanario:2011cs}. For loop integrals up to the box level,
we use Passarino-Veltman reduction~\cite{Passarino:1978jh}.
There we circumvent the explicit calculation of inverse Gram
determinants numerically by solving a system of linear equations, which
leads to a more stable behavior close to the critical regions.
For all loop contributions, the full virtual corrections
\begin{align}
M_V = \widetilde{M}_V + \ \frac{\alpha_S}{4 \pi} \ C_F \ \left( \frac{4
      \pi \mu^2}{Q^2} \right)^\epsilon  \ \Gamma{(1 + \epsilon)} \ \left[
    -\frac{2}{\epsilon^2} - \frac{3}{\epsilon} - 8 + \frac{4 \pi^2}{3}
  \right] \ M_B, 
\label{eq:MV}
\end{align}
factorize into a part proportional to the Born amplitude $M_B$ and an
infrared-finite remainder, $\widetilde{M}_V$. $Q$ denotes the partonic
center-of-mass energy, \ie{} the invariant mass of the photons and, for
the $Z\gamma\gamma$ case, the leptons. We note that for diagrams where
only a single vector boson is attached to the quark line (top right
diagram of Fig.~\ref{fig:feyn}), the virtual corrections completely
factorize to their corresponding Born matrix element, \ie{} they do not
contribute to $\widetilde{M}_V$.
The finite remainders for all virtual corrections to two-boson
amplitudes (top center diagram of Fig.~\ref{fig:feyn}) are called
``virtual-box'' and to all three vector bosons attached to the quark
line ``virtual-pentagons'' in the following. They were computed with the
method described in Ref.~\cite{Campanario:2011cs}.
To test the gauge invariance of the total $Z\gamma\gamma$ amplitude, we
write the polarization vector of the $Z$ as 
$\epsilon_Z^\mu = x_Z q_Z^\mu + \tilde\epsilon_Z^\mu$, where the last
part simply denotes the remaining piece. Due to the structure of the
amplitude, we cannot shift any contributions from pentagons to boxes, as
has been done in other triboson processes~\cite{Hankele:2007sb,
Campanario:2008yg, Bozzi:2009ig, Bozzi:2010sj}. Still, individual
pentagon amplitudes change significantly depending on the value of
$x_Z$. We have verified that their sum nevertheless stays unchanged.
%The standard value of $x_Z=0$ thereby gives the smallest cancellation
%between different parts.

Isolated photons in the final state require particular care for the real
emission part. Collinear emission of a photon from a massless quark
leads to additional infrared divergences. Applying a naive separation
cut between these two particles is not possible, however, since this would
restrict the phase-space of the final-state parton and therefore spoil
the cancellation of divergences between virtual and real part. In this
article, we solve the problem by imposing a particularly crafted
cut~\cite{Frixione:1998jh} between final-state parton (quark or gluon)
$i$ and photon $\gamma_j$
\begin{equation}
p_{T_i} \le p_{T_{\gamma_j}} \frac{1-\cos R_{\gamma_j i}}{1-\cos\delta_0} 
\text{\quad or\quad} R_{\gamma_j i} > \delta_0 \ ,
\label{eq:frixione}
\end{equation}
where $j$ runs over all final-state photons and $\delta_0$ is a fixed
separation parameter. This cut allows the parton to be arbitrarily close
to the photon, as long as its momentum vanishes simultaneously. This
procedure avoids electroweak divergences, keeping at the same time 
the QCD pole of the real part intact.

We have performed several cross checks to verify that our calculation is
correct. First, all tree-level matrix elements have been compared
against corresponding code generated by
\program{MadGraph}~\cite{Stelzer:1994ta}. We find an agreement at the
level of the machine precision, \ie{} at least 14 significant digits.
Furthermore, we have compared the integrated cross sections of all
leading-order processes as well as those with an additional jet against
\program{MadEvent} and \program{Sherpa}~\cite{Sherpa:09}. For the triple
photon process, internal technical cuts in these programs make the
comparison difficult when we impose rather loose cuts on the photons.
Therefore, we have performed an additional cross check with
\program{FeynArts}~\cite{FA}/\program{FormCalc}~\cite{FC}/\program{HadCalc}~\cite{HC}
for this instance. In all cases, we find an excellent agreement between
the codes below the per mill level, compatible with the integration
error. Also, the implementation of the Catani-Seymour subtraction scheme
has been checked. We have tested that the large contributions from the
true real-emission diagrams are canceled by the corresponding dipole terms
once we approach the soft or collinear regions. Additionally, we have
verified that finite contributions appearing in the subtraction scheme
can be shifted between the virtual and the real part without affecting
the total result. Furthermore, for the virtual corrections we employ a
gauge test for each phase-space point.

\section{Numerical Results}
\label{sec:numeric}

We obtain numerical results from an NLO Monte Carlo program based on the
structure of the \program{VBFNLO} program package~\cite{Arnold:2008rz}. In the
electroweak sector, we take the masses of the $W$ and $Z$ boson as well
as the Fermi constant as input. We then use tree-level relations to
calculate the weak mixing angle and the electromagnetic coupling from
these. As numerical values we hence use
\begin{align}
&m_W = 80.398 \ \mathrm{GeV} && \sin^2{(\theta_W)} = 0.22264 \nonumber\\ 
&m_Z = 91.1876 \ \mathrm{GeV} && \alpha^{-1} = 132.3407 \nonumber\\
&G_F = 1.16637 \cdot 10^{-5} \ \mathrm{GeV}^{-2}  
\; . &&  \label{eq:ew}
\end{align}
We do not consider any effects arising from top quarks. All other quarks
are considered to be massless. 
%CKM matrix does not appear.
As parton distribution functions we use the CTEQ6L1 set at
LO~\cite{Pumplin:2002vw} and the CT10 set with $\alpha_S(m_Z)=0.118$ at
NLO~\cite{Lai:2010vv}. For both sets, we use five active flavors as
initial-state quarks and in the renormalization group running of the
strong coupling constant.
As central value for the factorization and renormalization scale, we take
the invariant mass of all uncolored particles, \ie{}
\begin{align} \label{eq:scale}
\mu_F = \mu_R = \mu_0 =  
\begin{cases}
m_{Z\gamma\gamma} \equiv
\sqrt{(p_{\ell_1/\nu_1} + p_{\bar{\ell}_1/\bar{\nu}_1} 
       + p_{\gamma_1} + p_{\gamma_2})^2} & \text{for $Z\gamma\gamma$ and} \\
m_{\gamma\gamma\gamma} \equiv
\sqrt{(p_{\gamma_1}
       + p_{\gamma_2} + p_{\gamma_3})^2} & \text{for $\gamma\gamma\gamma$.} 
\end{cases}
\end{align}

To take into account typical requirements of the experimental detectors,
we impose the following set of minimal cuts on the transverse momentum
and rapidity of the final-state charged leptons and photons
\begin{align}
p_{T_{\gamma}} &> 20 \ \mathrm{GeV} &
p_{T_{\ell}} &> 20 \ \mathrm{GeV} &
|y_{\gamma}| &< 2.5 &
|y_{\ell}| &< 2.5 \ .
\label{eq:cuts1}
\end{align}
Furthermore, photons, charged leptons and jets need to be
well-separated in phase space, so they can be identified as separate
objects in the detector. Also, divergences from collinear photons need
to be eliminated by cuts. Hence, we use additionally
\begin{align}
R_{\ell\ell} &> 0.3  &
R_{\gamma\gamma} &> 0.4  &
R_{\ell\gamma} &> 0.4  &
R_{j \ell} &> 0.4  &
R_{j \gamma} &> 0.7 &
m_{\ell^+\ell^-} &> 15 \ \mathrm{GeV} \ ,
\label{eq:cuts2}
\end{align}
where the last requirement removes the singularity from a virtual photon
splitting into a pair of oppositely charged leptons, when the invariant
mass of the lepton pair becomes small. For the purpose of these cuts, a
jet is defined as a final-state gluon or quark with $p_{T_j} > 30$ GeV
and $|y_{j}| < 4.5$. Furthermore, we impose the restriction defined in
Eq.~(\ref{eq:frixione}) with the parameter $\delta_0 = 0.7$, which
eliminates electroweak divergences between final-state quarks or gluons
and photons collinear to them.

\begin{table}
  \begin{center}
    \begin{tabular*}{0.99\textwidth}{@{\extracolsep{\fill}}|l|cc|cc|cc|}
      \hline
      LHC 
       &\multicolumn{1}{c}{LO [fb]} &
       &\multicolumn{1}{c}{NLO [fb]} &
       & K-factor & \\ \hline
      \multicolumn{1}{|l|}{$~\sigma("Z\gamma\gamma" \to e^+ e^- \gamma \gamma)$} 
       &\multicolumn{2}{c|}{}
       &\multicolumn{2}{c|}{}
       &\multicolumn{2}{c|}{}\\
      $~~p_{T{\gamma(\ell)}}> 20(20)$ GeV 
       & \phantom{0}3.208\phantom{0}&
       & \phantom{0}4.877\phantom{0} &
       & 1.52 & \\ 
      $~~p_{T{\gamma(\ell)}}>30(20)$ GeV
       & \phantom{0}0.9987\phantom{} &
       & \phantom{0}1.565\phantom{0} &
       & 1.57 & \\ \hline
      \multicolumn{1}{|l|}{$~\sigma("Z\gamma\gamma" \to  \nu \bar \nu \gamma \gamma)$}
       &\multicolumn{2}{c|}{}
       &\multicolumn{2}{c|}{}
       &\multicolumn{2}{c|}{}\\
      $~~p_{T{\gamma}}>20$ GeV
       & \phantom{0}2.905\phantom{0} &
       & \phantom{0}4.510\phantom{0} &
       & 1.55 & \\       
      $~~p_{T{\gamma}}>30$ GeV       
       & \phantom{0}1.187\phantom{0} &
       & \phantom{0}1.906\phantom{0} &
       & 1.61 & \\ \hline
      \multicolumn{1}{|l|}{$~\sigma("\gamma \gamma\gamma" )$}
       &\multicolumn{2}{c|}{}
       &\multicolumn{2}{c|}{}
       &\multicolumn{2}{c|}{}\\
      $~~p_{T{\gamma}}>20$ GeV
       & \phantom{}22.17\phantom{00} &
       & \phantom{}58.57\phantom{00} &
       & 2.64 & \\       
      $~~p_{T{\gamma}}>30$ GeV       
       & \phantom{0}6.907\phantom{0} &
       & \phantom{}16.49\phantom{00} &
       & 2.39 & \\ \hline
    \end{tabular*}
    \caption[]{Total cross sections at the LHC for $pp \to
      Z\gamma\gamma+X$ with decays into charged leptons and neutrinos,
      and for $pp \to \gamma\gamma\gamma$. We show results for two sets
      of cuts at LO, NLO, and the associated K-factor. Relative
      statistical errors of the Monte Carlo are below $10^{-3}$.}
    \label{tab:LHC}
  \end{center}
\end{table}

\begin{table}
  \begin{center}
    \begin{tabular*}{0.99\textwidth}{@{\extracolsep{\fill}}|l|cc|cc|cc|}
      \hline
      Tevatron 
       &\multicolumn{1}{c|}{LO [fb]} & 
       &\multicolumn{1}{c|}{NLO [fb]} & 
       & K-factor & \\ \hline
      \multicolumn{1}{|l|}{$~\sigma("Z\gamma\gamma" \to e^+ e^- \gamma \gamma)$} 
       &\multicolumn{2}{c|}{}
       &\multicolumn{2}{c|}{}
       &\multicolumn{2}{c|}{}\\
      $~~p_{T{\gamma(\ell)}}> 10(10)$ GeV 
       & \phantom{}11.81\phantom{00} & 
       & \phantom{}16.37\phantom{00} & 
       & 1.39 & \\ 
      $~~p_{T{\gamma(\ell)}}>20(10)$ GeV
       & \phantom{0}1.568\phantom{0} & 
       & \phantom{0}2.226\phantom{0} & 
       & 1.42 & \\ \hline
      \multicolumn{1}{|l|}{$~\sigma("Z\gamma\gamma" \to  \nu \bar \nu \gamma \gamma)$}
       &\multicolumn{2}{c|}{}
       &\multicolumn{2}{c|}{}
       &\multicolumn{2}{c|}{}\\
      $~~p_{T\gamma}>10$ GeV
       & \phantom{0}2.421\phantom{0} & 
       & \phantom{0}3.534\phantom{0} & 
       & 1.46 & \\       
      $~~p_{T\gamma}>20$ GeV       
       & \phantom{0}0.5820\phantom{} & 
       & \phantom{0}0.8492\phantom{} & 
       & 1.46 & \\ \hline
      \multicolumn{1}{|l|}{$~\sigma("\gamma \gamma\gamma" )$}
       &\multicolumn{2}{c|}{}
       &\multicolumn{2}{c|}{}
       &\multicolumn{2}{c|}{}\\
      $~~p_{T\gamma}>10$ GeV
       & \phantom{}40.67\phantom{00} &
       & \phantom{}84.20\phantom{00} &
       & 2.07 & \\       
      $~~p_{T\gamma}>20$ GeV       
       & \phantom{0}6.008\phantom{0} &
       & \phantom{}10.28\phantom{00} &
       & 1.71 & \\ \hline
    \end{tabular*}
    \caption[]{Total cross sections at the Tevatron for $pp \to
      Z\gamma\gamma+X$ with decays into charged leptons and neutrinos,
      and for $pp \to \gamma\gamma\gamma$. We show results for two sets
      of cuts at LO, NLO, and the associated K-factor. Relative
      statistical errors of the Monte Carlo are below $10^{-3}$.}
    \label{tab:Tevatron}
  \end{center}
\end{table}

In Table~\ref{tab:LHC}, we present results for the integrated cross
section of $Z\gamma\gamma$ and $\gamma\gamma\gamma$ production at the
LHC with a center-of-mass energy of $14$~TeV at the central scale
$\mu_0$. Besides our standard set
of cuts in Eqs.~(\ref{eq:cuts1}) and~(\ref{eq:cuts2}), we also show
results with a modified cut on the transverse momentum of the photon of
30~GeV.  The results for the Tevatron with its center-of-mass energy of
$1.96$~TeV are denoted in Table~\ref{tab:Tevatron}. Here, we have reduced
the cut on ${p_T}_\ell$ to 10~GeV, and besides the standard value of
20~GeV also give results for ${p_T}_\gamma$ larger than 10~GeV. In all
cases, we consider only a single generation of charged leptons or
neutrinos in the final state.

\begin{figure}[htbp!]
\includegraphics[scale=1,clip]{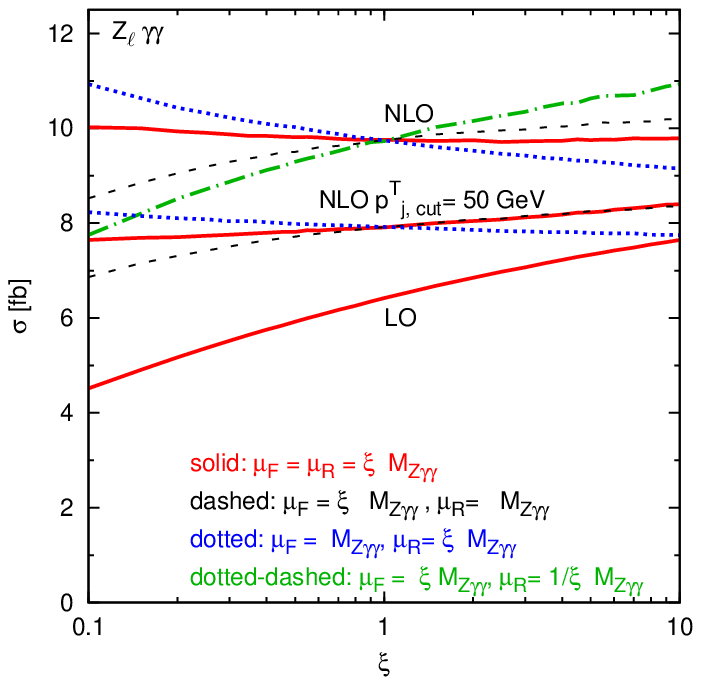}
\includegraphics[scale=1,clip]{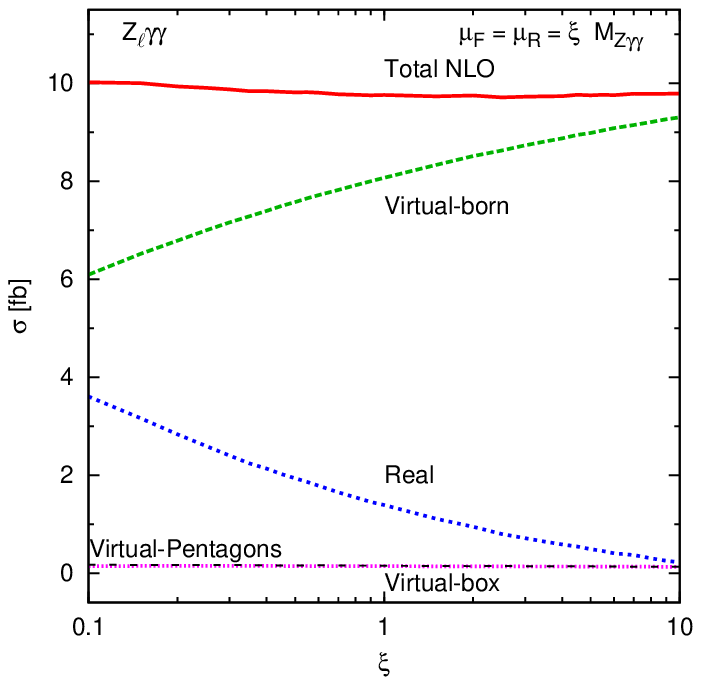}
\caption[]{
{\it Left:} Scale dependence of the total LHC cross section for 
  $p p \to  Z \gamma\gamma + X \to \ell^+ \ell^- \gamma\gamma + X$ 
  at LO and NLO within the cuts of Eqs.~(\ref{eq:cuts1}, \ref{eq:cuts2}).
  The factorization and renormalization scales are together, inversely or
  independently varied in the range from $0.1 \cdot \mu_0$ to $10 \cdot
  \mu_0$. 
  We show NLO curves without and including an additional veto on
  ${p_T}_j$ of 50 GeV.
{\it Right:} Same as in the left panel without jet veto, 
  but for the different NLO
  contributions at $\mu_F=\mu_R=\xi\mu_0$ with $\mu_0 =
  m_{Z\gamma\gamma}$.}
\label{fig:scaleZAA}
\end{figure}
\begin{figure}[htbp!]
\includegraphics[scale=1,clip]{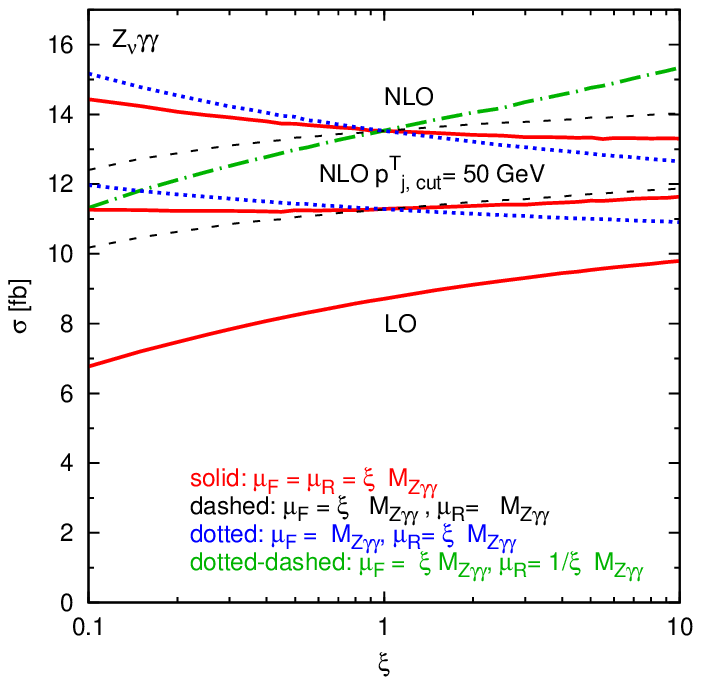}
\includegraphics[scale=1,clip]{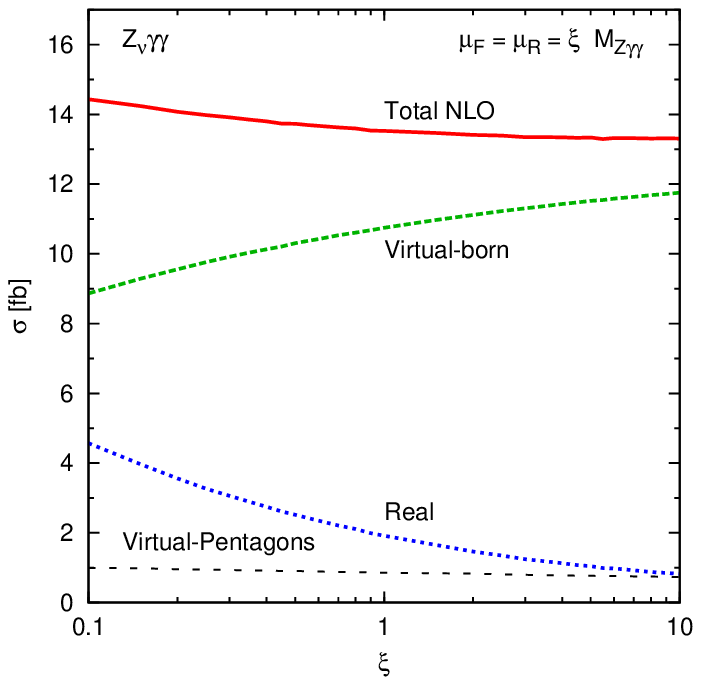}
\caption[]{
{\it Left:} Scale dependence of the total LHC cross section for 
  $p p \to  Z \gamma\gamma + X \to \nu \bar\nu \gamma\gamma + X$ 
  at LO and NLO within the cuts of Eqs.~(\ref{eq:cuts1}, \ref{eq:cuts2}).
  The factorization and renormalization scales are together, inversely or
  independently varied in the range from $0.1 \cdot \mu_0$ to $10 \cdot
  \mu_0$. 
  We show NLO curves without and including an additional veto on
  ${p_T}_j$ of 50 GeV.
{\it Right:} Same as in the left panel without jet veto, but for the
  different NLO contributions at $\mu_F=\mu_R=\xi\mu_0$ with $\mu_0 =
  m_{Z\gamma\gamma}$.}
\label{fig:scaleZnAA}
\end{figure}
\begin{figure}[htbp!]
\includegraphics[scale=1,clip]{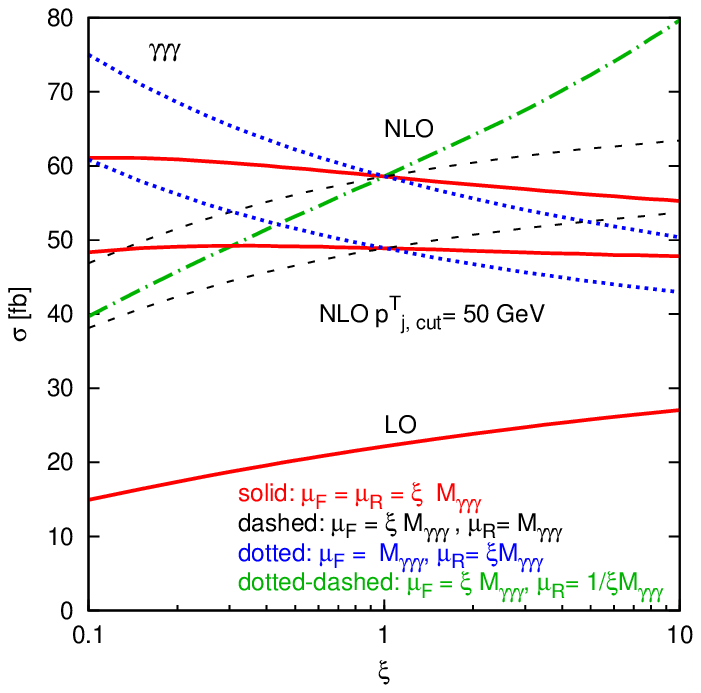}
\includegraphics[scale=1,clip]{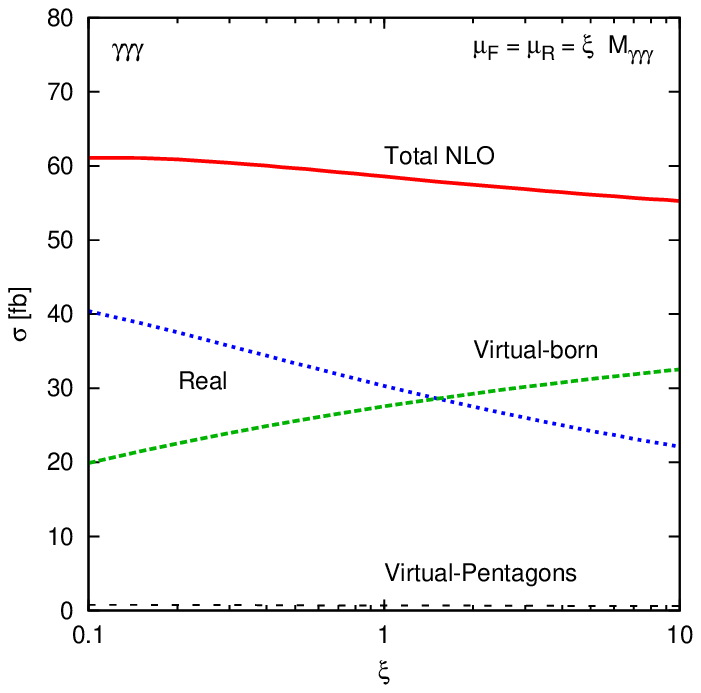}
\caption[]{
{\it Left:} Scale dependence of the total LHC cross section for 
  $p p \to  \gamma \gamma\gamma + X $ 
  at LO and NLO within the cuts of Eqs.~(\ref{eq:cuts1}, \ref{eq:cuts2}).
  The factorization and renormalization scales are together, inversely or
  independently varied in the range from $0.1 \cdot \mu_0$ to $10 \cdot
  \mu_0$. 
  We show NLO curves without and including an additional veto on
  ${p_T}_j$ of 50 GeV.
{\it Right:} Same as in the left panel without jet veto, but for the
  different NLO contributions at $\mu_F=\mu_R=\xi\mu_0$ with $\mu_0 =
  m_{\gamma\gamma\gamma}$.}
\label{fig:scaleAAA}
\end{figure}
From hereon, we will restrict ourselves to results for the LHC.
Additionally, we now sum over both electrons and muons in
$Z_\ell\gamma\gamma$ and over all three neutrino generations in
$Z_\nu\gamma\gamma$.
In Figs.~\ref{fig:scaleZAA} and~\ref{fig:scaleZnAA}, we show the
dependence of the cross section on the factorization and renormalization
scale for the $Z\gamma\gamma$ process with decays into charged leptons
and neutrinos, respectively. We vary both scales together,
independently and inversely, such that when one scale is increased, the other
one decreases by the same factor, in the range
\begin{equation}
\mu_F, \mu_R = \xi \cdot \mu_0 \quad (0.1 < \xi < 10) \ ,
\end{equation}
where the central scale $\mu_0$ has been given in Eq.~(\ref{eq:scale}).
The corresponding plot for the $\gamma\gamma\gamma$ process is displayed
in Fig.~\ref{fig:scaleAAA}. In all cases, we also show results where we
have applied an additional veto on jets with ${p_T}_j > 50 \text{ GeV}$.
This results in smaller renormalization scale variations, but the
principal shapes of the different curves are not affected.
We observe that for all processes, the NLO corrections are much larger
than estimated by a scale variation of the LO result. The respective
K-factors at the central scale are 1.52 and 1.55 for $Z\gamma\gamma$
decaying into charged leptons and into neutrinos, and 2.64 for
$\gamma\gamma\gamma$. 
We observe that the NLO cross section typically rises with the
factorization scale and decreases with the renormalization scale. This
leads to cancellations for the combined scale dependence which can yield
a very flat behavior, as can be seen for example in
Fig.~\ref{fig:scaleZAA}. In contrast, when we vary the two scales
inversely, the variations add up and lead to a rather significant
dependence. This almost flat behavior is therefore accidental, and
should not be considered as a very small remaining scale uncertainty. 

On the right-hand side of Figs.~\ref{fig:scaleZAA}, \ref{fig:scaleZnAA}
and~\ref{fig:scaleAAA}, we show the individual contributions to the
unvetoed NLO cross section for the different processes as function of
the combined factorization and renormalization scale. Both the
real-emission part, which contains the true real-emission contribution,
the dipole terms from the Catani-Seymour subtraction scheme and the
finite collinear terms, as well as the virtual part proportional to the
Born matrix element exhibit a strong scale dependence, which partly
cancels in the sum. For the $Z\gamma\gamma$ process, the bulk of the
contribution is given by the virtual part, while for
$\gamma\gamma\gamma$ it is evenly distributed between the two. This is
due to the higher average partonic center-of-mass energy for
$Z\gamma\gamma$, which reduces the initial-state gluon parts due to the
steep fall-off of the gluon pdfs with growing energy. The pentagon
finite virtual remainders, ``virtual-pentagons'', defined in
Eq.~(\ref{eq:MV}), are small in all cases. Finite box virtual
remainders, ``virtual-box'', exist only for the
$Z_\ell\gamma\gamma$ process and they are small as well.

\begin{figure}[htbp!]
\includegraphics[scale=1,clip]{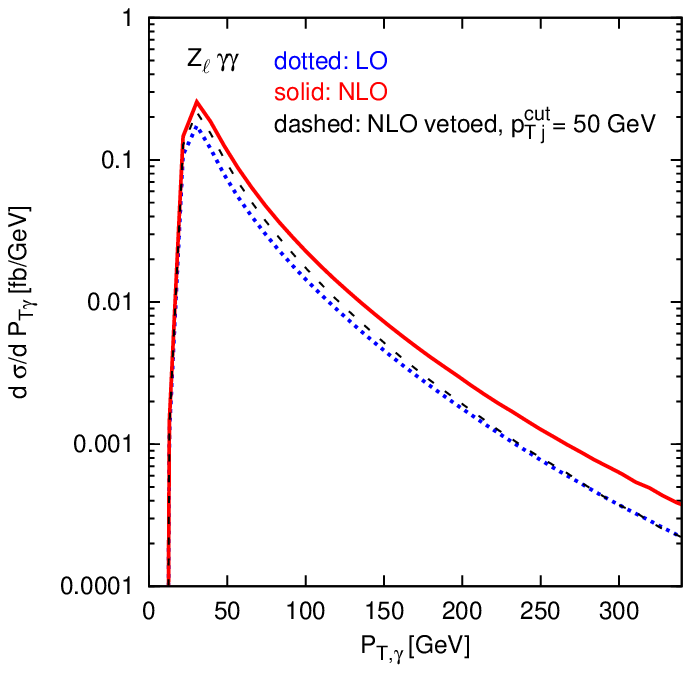}
\includegraphics[scale=1,clip]{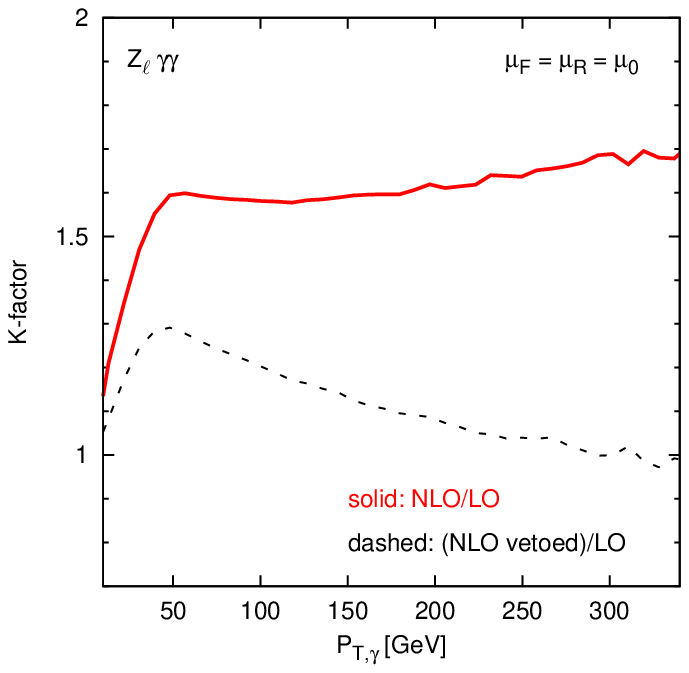}
\caption[]{
{\it Left:} Transverse-momentum distribution of the photon with largest
transverse momentum in $Z_\ell\gamma\gamma$ production with $Z$ decaying
into charged leptons for the LHC. We show LO (dotted blue line)
and NLO cross sections without (solid red) and including (dashed black)
a jet veto of 50 GeV, using the cuts of Eqs.~(\ref{eq:cuts1},
\ref{eq:cuts2}).
{\it Right:} Associated K-factor as defined in Eq.(\ref{eq:kfactor})
without (solid red) and including (dashed black) the jet veto.}
\label{fig:ZAA_ptAhard}
\end{figure}
\begin{figure}[htbp!]
\includegraphics[scale=1,clip]{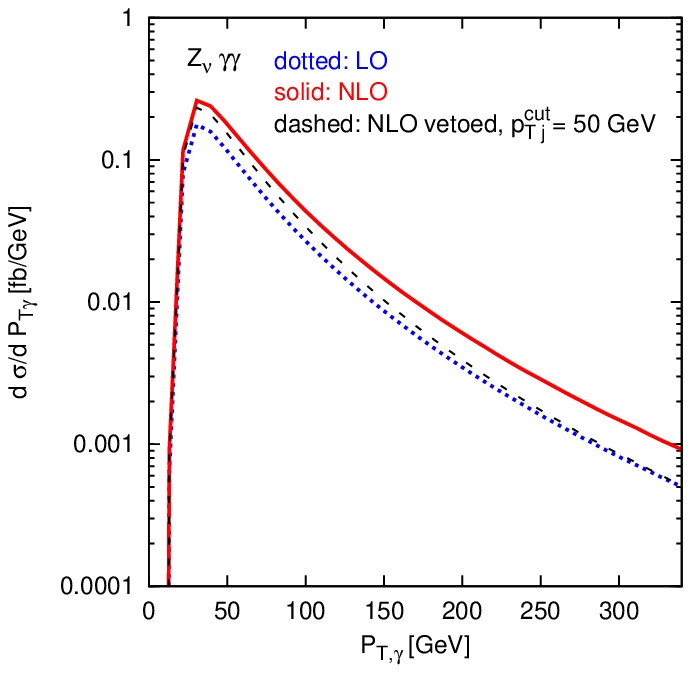}
\includegraphics[scale=1,clip]{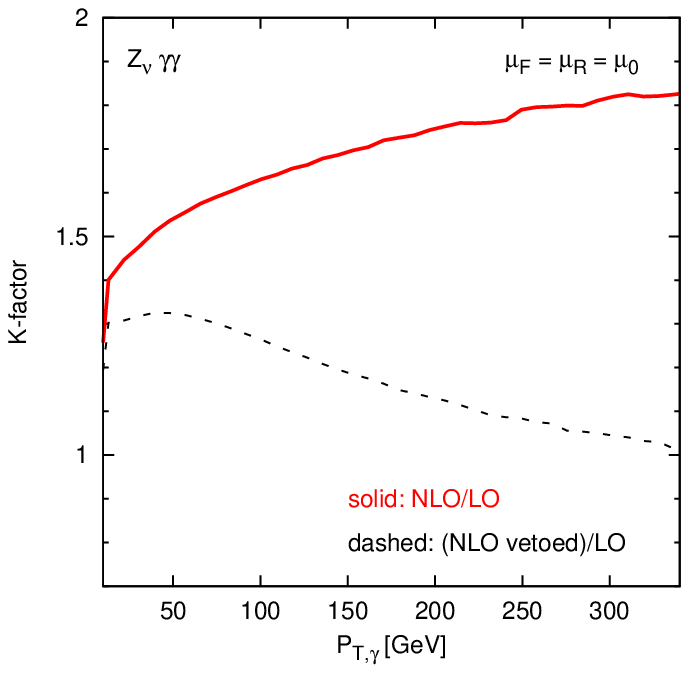}
\caption[]{
{\it Left:} Transverse-momentum distribution of the photon with largest
transverse momentum in $Z_\nu\gamma\gamma$ production with $Z$ decaying
into neutrinos for the LHC. We show LO (dotted blue line) and NLO
cross sections without (solid red) and including (dashed black) a jet
veto of 50 GeV, using the cuts of Eqs.~(\ref{eq:cuts1}, \ref{eq:cuts2}).
{\it Right:} Associated K-factor as defined in Eq.(\ref{eq:kfactor})
without (solid red) and including (dashed black) the jet veto.}
\label{fig:ZnAA_ptAhard}
\end{figure}
\begin{figure}[htbp!]
\includegraphics[scale=1,clip]{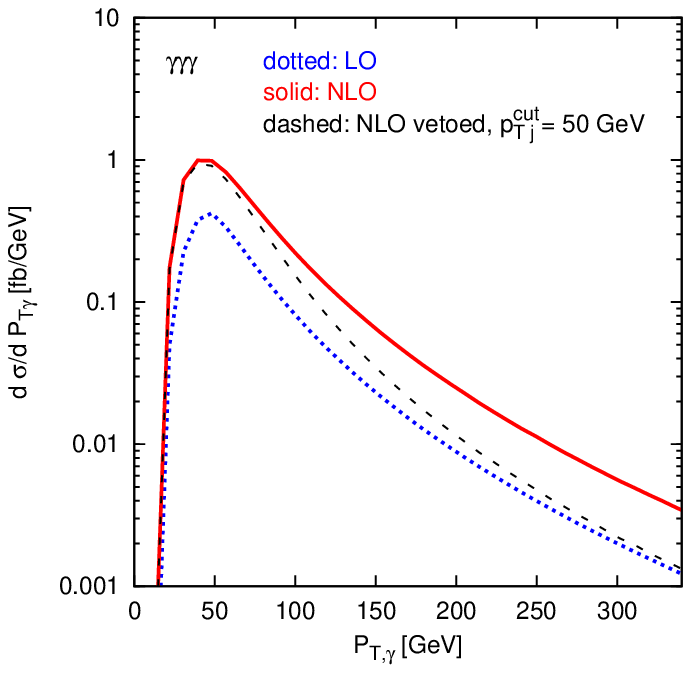}
\includegraphics[scale=1,clip]{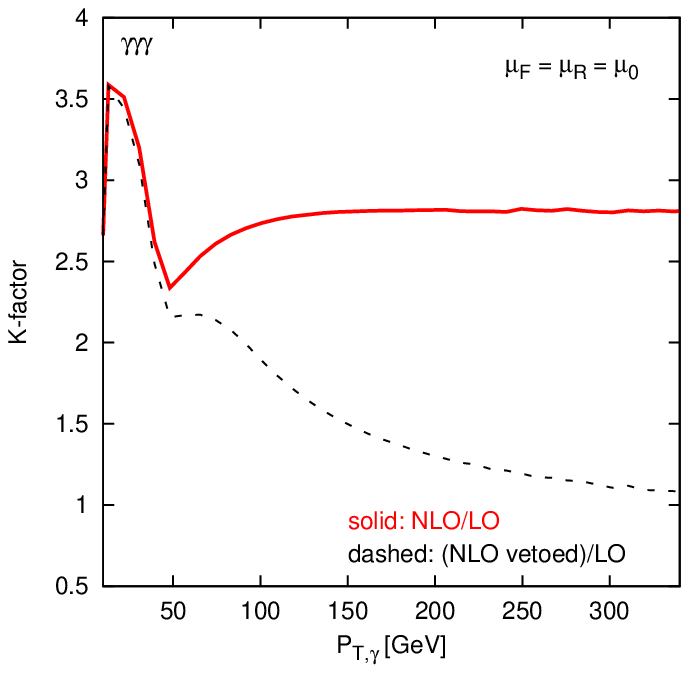}
\caption[]{
{\it Left:} Transverse-momentum distribution of the photon with largest
transverse momentum in $\gamma\gamma\gamma$ production for the LHC. We
show LO (dotted blue line) and NLO cross sections without (solid red)
and including (dashed black) a jet veto of 50 GeV, using the cuts of
Eqs.~(\ref{eq:cuts1}, \ref{eq:cuts2}).  
{\it Right:} Associated K-factor
as defined in Eq.(\ref{eq:kfactor}) without (solid red) and including
(dashed black) the jet veto.}
\label{fig:AAA_ptAhard}
\end{figure}
\begin{figure}[htbp!]
\includegraphics[scale=1,clip]{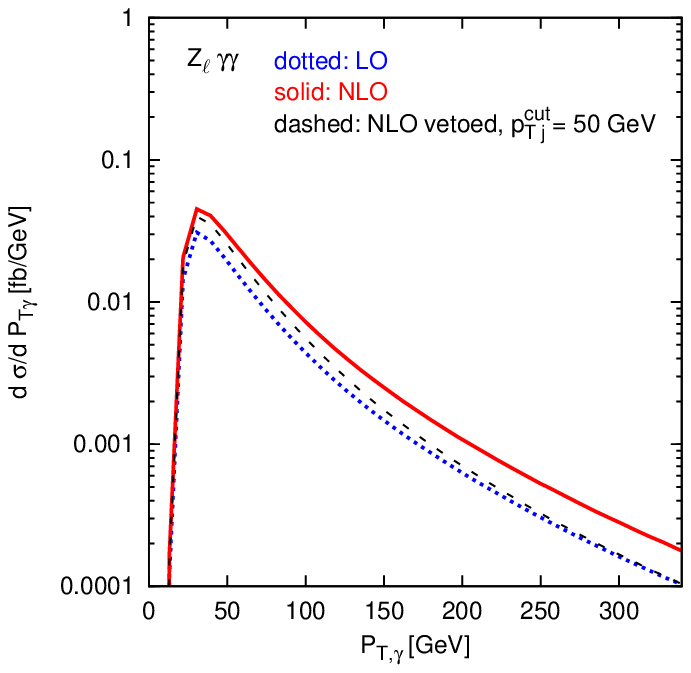}
\includegraphics[scale=1,clip]{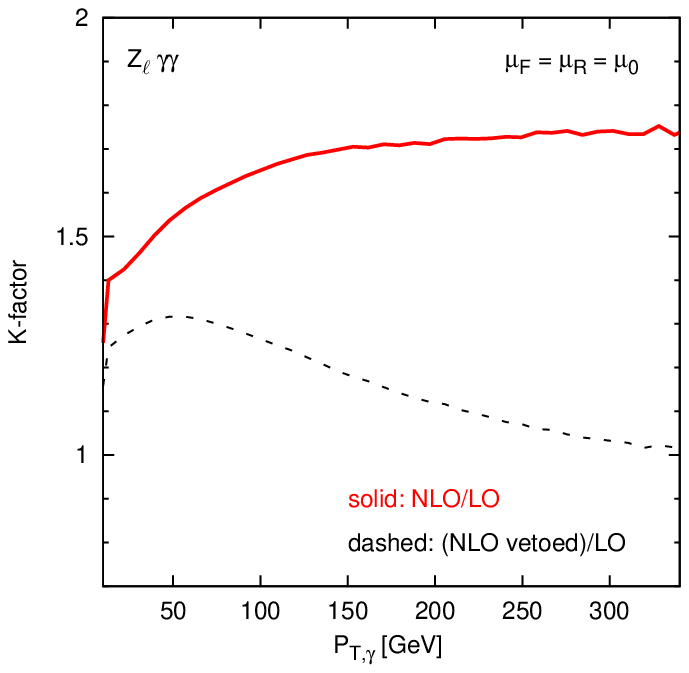}
\caption[]{
{\it Left:} Transverse-momentum distribution of the photon with largest
transverse momentum in $Z_\ell\gamma\gamma$ production with $Z$ decaying
into charged leptons for the LHC. We show LO (dotted blue line)
and NLO cross sections without (solid red) and including (dashed black)
a jet veto of 50 GeV, using the cuts of Eqs.~(\ref{eq:cuts1},
\ref{eq:cuts2}) and the additional cut on $m_{\ell\ell}$ given in
Eq.~(\ref{eq:cutmll}).
{\it Right:} Associated K-factor as defined in Eq.(\ref{eq:kfactor})
without (solid red) and including (dashed black) the jet veto.}
\label{fig:ZAAcut_ptAhard}
\end{figure}
\begin{figure}[htbp!]
\includegraphics[scale=1,clip]{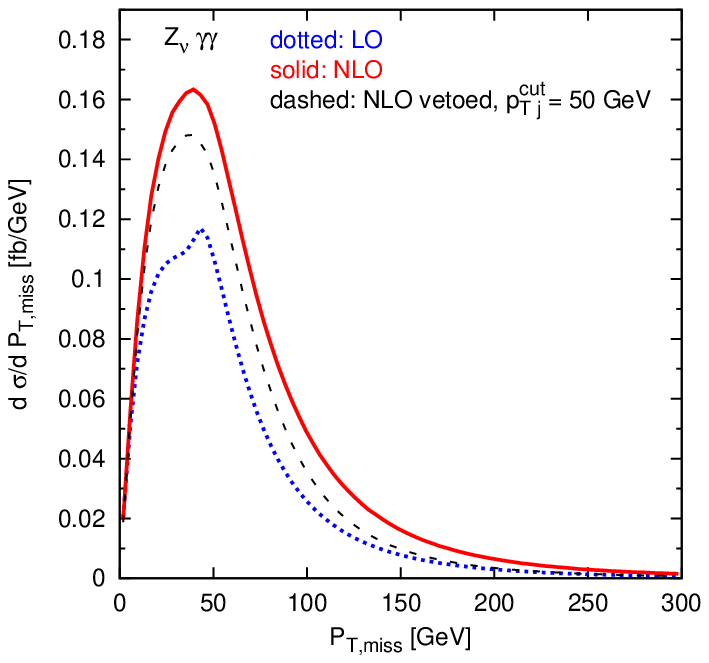}
\includegraphics[scale=1,clip]{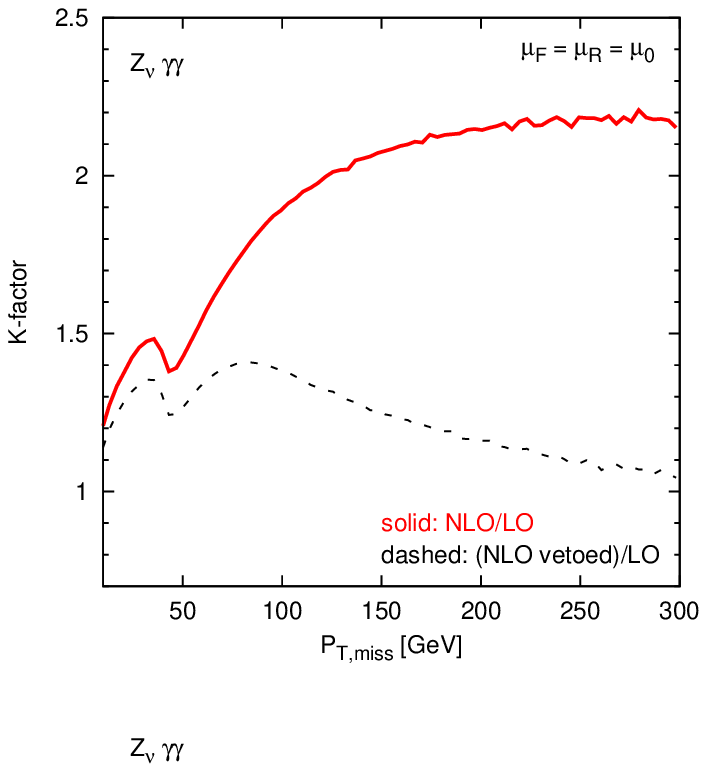}
\caption[]{
{\it Left:} Distribution of missing transverse momentum in
$Z_\nu\gamma\gamma$ production with $Z$ decaying into neutrinos for
the LHC. We show LO (dotted blue line) and NLO cross sections without
(solid red) and including (dashed black) a jet veto of 50 GeV, using the
cuts of Eqs.~(\ref{eq:cuts1}, \ref{eq:cuts2}).
{\it Right:} Associated K-factor as defined in Eq.(\ref{eq:kfactor})
without (solid red) and including (dashed black) the jet veto.}
\label{fig:ZnAA_ptmiss}
\end{figure}
\begin{figure}[htbp!]
\includegraphics[scale=1,clip]{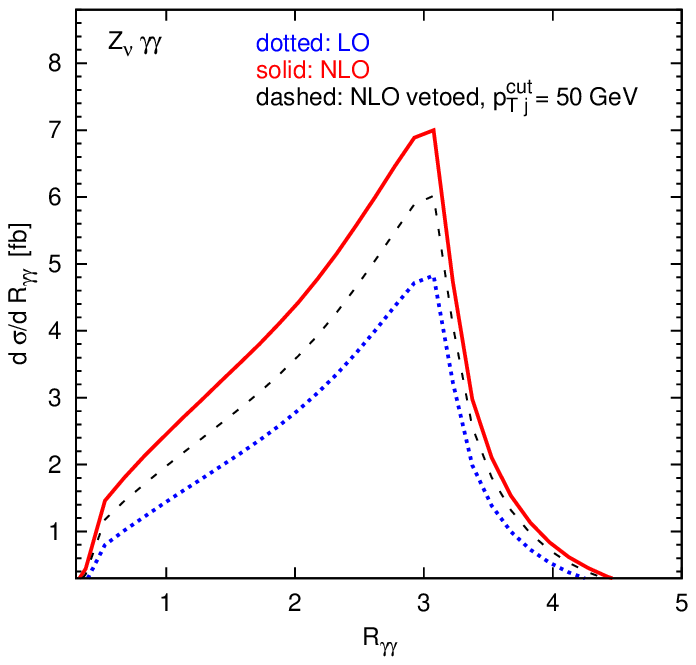}
\includegraphics[scale=1,clip]{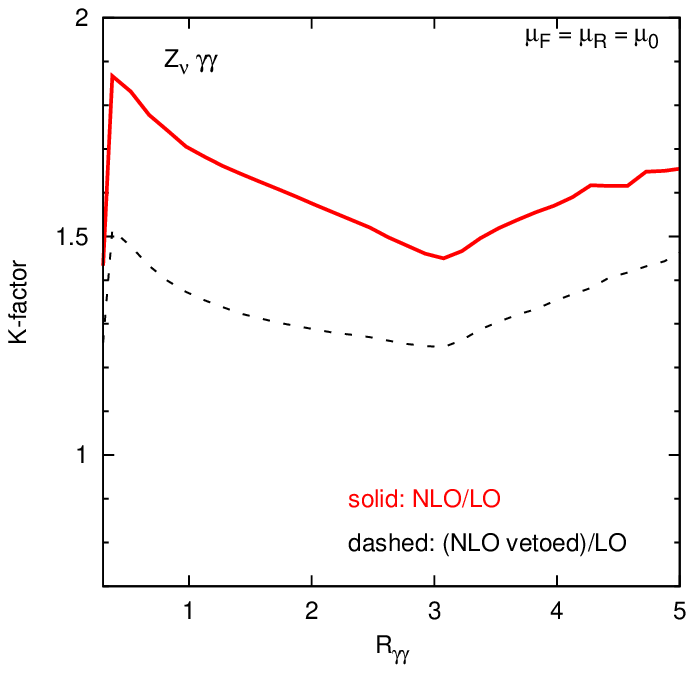}
\caption[]{
{\it Left:} Separation of the two photons in $Z_\nu\gamma\gamma$
production with $Z$ decaying into neutrinos for the LHC. We show LO
(dotted blue line) and NLO cross sections without (solid red) and
including (dashed black) a jet veto of 50 GeV, using the cuts of
Eqs.~(\ref{eq:cuts1}, \ref{eq:cuts2}).
{\it Right:} Associated K-factor as defined in Eq.(\ref{eq:kfactor})
without (solid red) and including (dashed black) the jet veto.}
\label{fig:ZnAA_RAA}
\end{figure}

Finally, we look at several differential distributions for all
considered processes. In each figure, we show at the left-hand side the
differential cross section both at LO and NLO. The latter we plot
without and including an additional veto on jets with ${p_T}_j > 50
\text{ GeV}$. On the right-hand side, we depict the differential K-factor
for both the unvetoed results and including the jet veto.
It is defined as 
\begin{align}
K = \frac{d \sigma^{NLO}/ dx}{d \sigma^{LO} /dx} \ ,
\label{eq:kfactor}
\end{align}
where $x$ denotes the considered observable.

In Figs.~\ref{fig:ZAA_ptAhard}, \ref{fig:ZnAA_ptAhard} and
\ref{fig:AAA_ptAhard}, we present the transverse-momentum distributions
of the photon with the largest transverse momentum for
$Z_\ell\gamma\gamma$, $Z_\nu\gamma\gamma$ and $\gamma\gamma\gamma$,
respectively. 
For each of the different processes, the K-factor for the differential
distribution is not constant. Rather, we observe for the
$Z\gamma\gamma$ processes a significant rise with larger transverse
momenta. At small transverse momenta, the K-factor is close to the
integrated results, as this is where the bulk of the cross section lies.
Once we apply the additional jet veto, the K-factor instead drops to
unity as the transverse momentum of the photon increases. Hence, the
large differential K-factors are caused by events where the leptonic
system recoils against the additional jet. For tri-photon production
(Fig.~\ref{fig:AAA_ptAhard}), we observe a substantially larger K-factor
for very small transverse photon momenta, where the recoil against the
jet helps to fulfill the transverse-momentum cut, but is not yet large
enough to trigger the possible jet veto. Once we go to large values, the
K-factor is almost constant without jet veto, and gradually drops to one
including it.

Furthermore, we show this distribution in Fig.~\ref{fig:ZAAcut_ptAhard}
for $Z_\ell\gamma\gamma$, where we have applied the additional cut 
\begin{equation}
m_Z - 2 \Gamma_Z < m_{\ell\ell} < m_Z + 2 \Gamma_Z  
\label{eq:cutmll}
\end{equation}
on the invariant mass $m_{\ell\ell}$ of the lepton pair. This cut allows
us to restrict the process in such a way, that the photons are
predominantly radiated off the quark lines (top left diagram of
Fig.~\ref{fig:feyn}), which is the only possibility for $Z$ decays into
neutrinos. Comparing this figure with Fig.~\ref{fig:ZnAA_ptAhard}, we
see that the shape of the plots both for the cross sections as well as
for the K-factors is almost identical. We have also checked several
other distributions. Up to a global normalization constant, there are
only small and negligible differences between the two. Therefore, by
using this cut it is possible to extrapolate the differential rate with
$Z$ decays into neutrinos from the charged-lepton rate. 

In Figs.~\ref{fig:ZnAA_ptmiss} and~\ref{fig:ZnAA_RAA}, we present two
more distributions for the $Z\gamma\gamma$ process with decay into
neutrinos. The distribution of missing transverse momentum originating
from the neutrino pair is shown in Fig.~\ref{fig:ZnAA_ptmiss}. At LO, we
see a broad distribution superimposed with a peak above $2\,
{p_T}_{\gamma,\min} = 40\text{ GeV}$, which is smeared out at NLO. This
gives rise to the structure at low momenta in the K-factor, which we
observe in the figure. For large missing momentum, we see again a
significantly larger K-factor without any veto. With an additional jet
veto of 50 GeV the K-factor drops to unity in this region.

Finally, in Fig.~\ref{fig:ZnAA_RAA} we show the separation in the
rapidity--azimuthal-angle plane between the two photons,
$R_{\gamma\gamma}$, for $Z_\nu\gamma\gamma$ . Also here, we observe a
large variation of the differential K-factor with the value of the
$R_{\gamma\gamma}$ separation. Both curves without and including the
additional jet veto exhibit a similar shape.

Therefore, it is necessary to calculate fully differential NLO cross
sections to obtain the correct shape of the distributions. A simple
rescaling of the LO result with the integrated K-factor would give large
deviations.

\section{Conclusions}
\label{sec:concl}

We have calculated the NLO QCD corrections to the processes
$pp, p\bar{p} \to Z\gamma\gamma + X$ with decays of the $Z$ boson into
both charged leptons and neutrinos and including all off-shell effects,
as well as to $pp, p\bar{p} \to \gamma\gamma\gamma + X$. These processes
appear as important background in searches for new physics, for example
in supersymmetric GMSB models or the latter in technicolor models.

In all cases, we obtain large corrections to the leading-order cross
section. The typical size of the integrated K-factors is around
1.5 for $Z\gamma\gamma$ and 2.6 for triple-photon production. This
strongly exceeds the expectations from LO scale variation, which is at
the level of 10\% to 16\%. For an estimate of the remaining scale
dependence, we vary the factorization and renormalization scale by a
factor two around the central scale 
$\mu_0 = m_{Z\gamma\gamma/\gamma\gamma\gamma}$ such that the product of
the two scales stays constant, \ie{} when one scale takes the factor
two, the other has factor $\tfrac12$ and vice versa. This gives the
largest differences between $\xi=2$ and $\xi=\tfrac12$ with changes of
9.0\%, 7.5\% and 18.6\% for $Z_\ell\gamma\gamma$, $Z_\nu\gamma\gamma$
and $\gamma\gamma\gamma$ at NLO, respectively. For a joint variation,
the difference is below 4\% for all processes due to accidental
cancellations.

The size of the NLO corrections exhibits a strong dependence on the
observable and on different regions of phase-space. We also observe that
for the processes considered in this article an additional jet veto does
not flatten the differential K-factors. Hence, it is necessary to have a
dedicated fully-exclusive NLO parton Monte Carlo program available.
We have also studied, whether a measurement of the charged-lepton decay mode
in $Z\gamma\gamma$ will allow us to predict the neutrino case. Here, we
find that this works well with an additional cut, which restricts the
invariant mass of the lepton pair to the region close to the $Z$ pole.

With these processes, we have completed the NLO QCD calculations of triple
vector-boson production at hadron colliders including off-shell effects,
Higgs resonances and leptonic decays. All these processes will be
available in the next release of \program{VBFNLO}.

%%%%%%%%%%%%%%%%%%%%%%%% appendix %%%%%%%%%%%%%%%%%%%%%%%%%%%%%%%%%%%%%%%
%\appendix
%\section{}

%%%%%%%%%%%%%%%%%%%%%%%% acknowledgements %%%%%%%%%%%%%%%%%%%%%%%%%%%%%%%
\begin{acknowledgments}
This research was supported in part by the Deutsche
Forschungsgemeinschaft via the Sonderforschungsbereich/Transregio
SFB/TR-9 ``Computational Particle Physics'' and the Initiative and
Networking Fund of the Helmholtz Association, contract HA-101 
(``Physics at the Terascale''). 
The Feynman diagrams in this paper were drawn using Jaxodraw~\cite{axo}. 
\end{acknowledgments}

%%%%%%%%%%%%%%%%%%%%%%%% biblio %%%%%%%%%%%%%%%%%%%%%%%%%%%%%%%%%%%%%%%%%

\end{document}